\documentclass[a4paper,11pt]{article}
\pdfoutput=1 

\usepackage{jheppub} 

\usepackage[T1]{fontenc} 
\usepackage{breqn}
\usepackage{derivative}
\usepackage{amsmath,mathrsfs}
\usepackage{float}
\usepackage{subcaption}
\usepackage{appendix}
\usepackage[numbers,sort&compress]{natbib}
\usepackage{notoccite}

\title{\boldmath }

\title{Dilute axion stars converting to photons in the Milky Way's magnetic field}

\author[a]{A. Kyriazis}

\affiliation[a]{Department of Physics,\\ University of Florida , \\Gainesville, FL 32611, \\ United States }

\emailAdd{akyriazis@ufl.edu}

\abstract{In this paper we examine the possibility of dilute axion stars converting to photons in the weak, large-scale magnetic field of the Milky Way and show that they can resonate with the surrounding plasma and produce a sizable signal. We consider two possibilities for the plasma: free electrons and HII regions. In the former case, we argue that the frequency of the photons will be too small to be observed even by space-based radio telescopes. In the latter case, their frequency is larger, safely above the solar wind cut-off. We provide an estimate of the flux as a function of the decay constant and show that for $f_{a} < 5 \times 10^{11} \text{GeV}$, the signal will be above the radio emission of the solar system's planets and it could potentially be detected by the NCLE instrument which is on board the Chang'e-4 spacecraft. Finally, we  calculate the time scale of decay of the axion star and demonstrate that back-reaction can be neglected for all physically interesting values of the decay constant, while the minimum time scale of decay is in the order of a few hours. }

\begin{document} 
\maketitle
\flushbottom

\section{Introduction}
\label{sec:intro}
\par The axion, initially proposed as a solution to the strong CP problem \cite{Peccei,Weinberg}, is now one of the most well-motivated candidates of dark matter \cite{Preskill,Sikivie, Willy}. The axions are stable bosons, with large occupation numbers and can re-thermalize through their gravitational interactions forming a Bose-Einstein condensate (BEC) \cite{BEC1,BEC2}. Owing to the large occupation number of the ground state, the BEC condensate has been treated classically as a localised, coherently oscillating clump called an axion star, if the kinetic pressure is balanced by gravity and axiton or oscillon, if it is balanced by self-interactions. Generally, when an axion star is supported only by its self-interactions, such as the cosine potential, it is considered dense with radius $mR \sim 1$ \cite{axion_star1} and also decays through scalar radiation on a time-scale of $10^{3} m^{-1}$ \cite{decay,Wilczek}. 
\par
On the other hand, if both gravity and the leading term in the self-interactions are taken into account, the size is $mR \sim \frac{M_{pl}}{f_{a}}$, where $M_{pl}=\frac{1}{\sqrt{8 \pi G}}= 2.4 \times 10^{18}  \text{GeV}$ is the reduced Planck mass and $f_{a}$ is the axion decay constant \cite{Chavanis, Wilczek}. Since the axion decay constant ranges from $10^{9} \text{GeV} < f_{a} < M_{pl}$ in the case where the Peccei-Quinn symmetry is broken during inflation and $10^{9} \text{GeV}< f_{a} < 10^{11} \text{GeV}$ in the classic axion window \cite{axion_cosmology}, these axion stars can be quite large for a wide range of decay constants. In contrast to dense axion stars, they are also long-lived, which makes them cosmologically very interesting \cite{axion_star2}. However, it has been shown that dilute axion stars become increasingly unstable as $f_{a}$ approaches $M_{pl}$ \cite{planck_decay}, because their binding energy becomes large and relativistic contributions need to be taken into account. Therefore, we will limit our analysis in the range $10^{9} \text{GeV} < f_{a} < 10^{15}  \text{GeV}$.
\par The axion can also interact with electromagnetic fields and several venues have been proposed for their detection \cite{axion_detection}. Of particular interest is the Primakoff effect \cite{Primakoff}, which is the interaction of an axion with a magnetic field to produce a real photon. 
\par
The universe is abundant with magnetic fields and many astrophysical settings have been considered in the literature as possible "laboratories" where the axion to photon conversion could be detected. These range from pulsars \cite{axion_neutron, Bai, transient_radio, Radio_signals,Radio_line,FRB} and magnetars \cite{magnetar1,magnetar2,magnetar3}, to white dwarfs \cite{upper_limit_from_axion_photon,white_dwarfs,spectral_distortions}, to AGN's \cite{AGN1,AGN2}, to the galactic magnetic field \cite{turbulent,axion_dm_radio_tele,relat_axions_in_sky,axionic_radiation}. 
\par In this paper, we will consider the possibility of an axion star converting to photons in the Milky Way's magnetic field. The strength of the magnetic field is of the order $1 \mu \text{G}$ and its coherence length is of the order of the galactic scale \cite{B_field_in_milky}. We will confirm that the flux emitted from dense axion stars in this magnetic field is negligible \cite{Bai}. However, there is the possibility of resonant conversion of axion stars to photons, if they find themselves in some region with cold plasma. By resonance, we mean that the axion mass equals the plasma frequency. It has been shown that in that case, the emitted power scales as $(mR)^{6}$, enhancing it significantly for dilute axion stars \cite{dipole, time_dependent}. The end result of the main calculation of this paper will be an expression for the emitted spectral flux density of \textit{dilute} axion stars in some region with cold plasma. 
\par
When it comes to the plasma, the average electron density in the Milky Way is of the order of $n_{e} \sim 0.03 \text{cm}^{-3}$, which implies a plasma frequency of the order $\omega_{p}=\sqrt{\frac{4 \pi \alpha n_{e}}{m_{e}}} \sim 10^{-12} \text{eV}$ \cite{radio_astronomy}. In the case of diffuse nebulae consisting of ionized hydrogen, the electron density ranges from 100-1000 $\text{cm}^{-3}$, with plasma frequencies in the 100-200 kHz range. \cite{ISM}. Because of energy conservation, the plasma frequencies that we estimated above will be the frequencies of the monochromatic photons that are emitted from the axion star. In the former case, those electromagnetic waves will be blocked by the solar wind, but in the latter case, they are in principal detectable by current space based radio telescopes, such as the NCLE \cite{NCLE}.
\par
Since the QCD axion's mass and decay constant must satisfy the equation $m_{a} f_{a} \approx (10^{8} \text{eV})^{2}$, due to the range of decay constant we are considering, our analysis does not cover the QCD axion, so we will only consider Axion-Like Particles (ALP's), for which the axion mass and the decay constant are independent \cite{ALP}.
\par 
In addition, we will demonstrate that our neglect of back-reaction effects is valid for $f_{a} > 10^{7} \text{GeV}$ and therefore for the entirety of the parameter space we are investigating. Finally, we will place a lower bound in the decay time scale, which will be in the order of a few hours. 
\par To motivate our idea further, we can make an order of magnitude estimate of the axion star number in the Milky Way: let's assume that only 1\% of the Milky Way's dark matter mass, $10^{12} M_{\odot}$,is distributed in axion stars. A dilute axion star has a mass of the order $10 \frac{M_{pl} f_{a}}{m}$. For $m \sim 10^{-12} \text{eV}$ and $f_{a} \sim 10^{13} \text{GeV}$, a typical axion star mass is $10^{-4} M_{\odot}$. These should be distributed over the galactic halo, but since we are only interested in those in the galactic disk, their number is $10^{14} \frac{V_{disk}}{V_{halo}} \sim 10^{11}$, where we used a typical radius of the halo 30 kpc and a disk radius of 15 kpc and height of 300 pc \cite{ANS_binaries}. This is indeed a huge number which makes the study of their resonant conversion in the Milky Way's magnetic field an intriguing possibility. 
 
\par The paper is organised as follows: In section 2, we establish our formalism that describes a dilute axion star. In section 3, we review some properties of the Milky Way's magnetic field, its free electron and HII distribution and finally discuss some prospects of detection of the emitted radiation by radio telescopes. In section 4, we outline the derivation of the conversion of an axion star to photons in a constant magnetic field, in the presence of cold plasma, and apply it to the case of a dilute star, while we also give an estimate of the spectral flux density of photons that will arrive at Earth from such an event. In section 5 we estimate the decay time-scale of the axion star and we conclude in section 6 with some comments on future venues for research. 

\section{Dilute Axions stars}

\par In this paper we will mainly focus on dilute axion stars with gravitational as well as attractive self interactions. Also, we focus on ALP's for which the axion decay constant $f_{a}$ and the axion mass $m_{a}$ are independent from each other. The action for a scalar field $\phi$ which describes the axion star, coupled to a gravitational potential $\Phi$ with attractive $\lambda \phi^{4}$ interactions is \cite{linear_newt}:
\begin{equation}
    \label{eqn: action}
    S=-\int d^{4}x \sqrt{-g} \left( \frac{1}{2}g^{\mu \nu} \partial_{\mu}\phi \partial{\nu} \phi +\frac{1}{2} m^{2} \phi^{2} - \frac{\lambda}{4!} \phi^{4} \right)
\end{equation}
with metric:

\begin{equation}
    \label{eqn: metric}
    ds^{2}=-(1+2\Phi(\vec{x},t))dt^{2}+d\vec{x} \cdot d\vec{x}
\end{equation}

Throughout this paper we will assume that $\lambda=\mathcal{O}(1) \frac{m^{2}}{f_{a}^{2}}$ and we'll ignore the order 1 constant, since we are interested in order of magnitude estimates. In the non-relativistic limit:
\begin{equation}
    \label{non-relat}
    \phi=\frac{1}{\sqrt{2m}}(\psi e^{-i m t}+\psi^{\star}e^{i m t})
\end{equation}
the equation of motion is the Gross-Pitaevskii equation:
\begin{equation}
    \label{eqn: Gross-Pit}
    i \partial_{t} \psi=-\frac{1}{2m} \nabla^{2} \psi + m \Phi \psi - \frac{\lambda}{8 m^{2}} |\psi|^{2} \psi
\end{equation}
while the potential satisfies the usual Poisson equation:
\begin{equation}
    \label{eqn: Poisson}
    \nabla^{2} \Phi = 4 \pi G m |\psi|^{2}
\end{equation}
 A stationary, spherically symmetric solution to the above equations is given by $\psi= e^{-i \mu t} \chi (r)$, where $\mu$ can be considered as the chemical potential. As a side remark that will be useful later on, we observe that if we insert this Ansatz into equation \ref{non-relat}, we get:
 \begin{equation}
     \label{eqn: Ansatz}
     \phi= \sqrt{\frac{2}{m}} \chi(r) \cos(\omega t) 
 \end{equation}
 where we have identified the frequency $\omega$ with $\omega=\mu+m$ \cite{global_view}. Since we are interested in bound states, it holds that $0<\frac{\omega}{m}<1$. This matches the Ansatz of the scalar field $\phi$ that has been used by other authors to study non-relativistic axion stars \cite{Wilczek},\cite{time_dependent} \cite{dipole}, with the difference of the $\sqrt{\frac{2}{m}}$ factor in front.

\begin{figure}[t]
    \centering
    \includegraphics{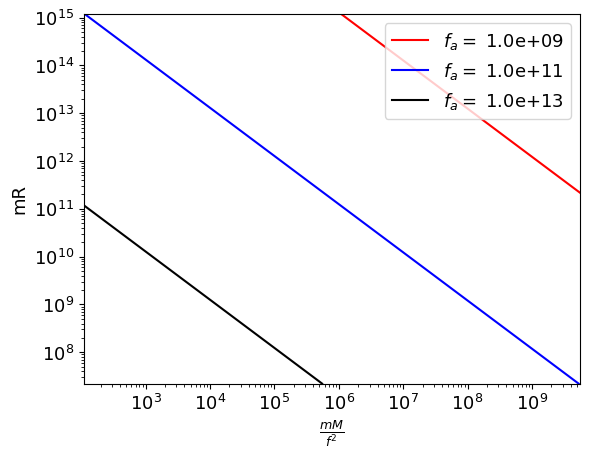}
    \caption{The mass-radius graph of dilute axion stars for three different values of $f_{a}$. Note that we have reverted back to the usual definitions of the dimensionless mass and radius that are found in the literature. The star becomes smaller as $f_{a} \rightarrow M_{pl} $.}
    \label{fig:mass_radius}
\end{figure}

\begin{figure}
    \centering
    \includegraphics{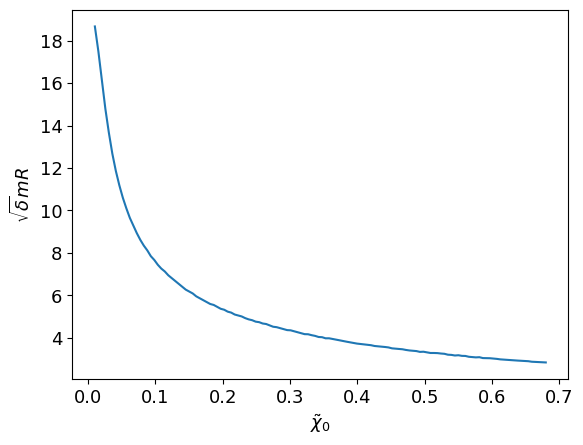}
    \caption{The dimensionless radius of the axion star as a function of its central amplitude}
    \label{fig:radius_ampl}
\end{figure}

 \par 
 To continue, we define the small parameter $\delta= \frac{4 m^{2}}{\lambda M_{pl}^{2}}$. Notice that if we plug in the value of $\lambda$ in terms of the axion mass and decay constant, we get $\delta=4 \frac{f_{a}^{2}}{M^{2}_{pl}}$, which is indeed small for a wide range of physically relevant decay constants \cite{axion_cosmology}. We rescale the wavefunction, the potential and the lengths to find the dimensionless forms of the above equations: 
\begin{equation}
    \label{eqn:rescale}
    \chi(r)=\sqrt{\frac{m}{4 \pi G}} \delta \tilde{\chi}(r), \hspace{0.4 cm} \vec{x}=\frac{\vec{\tilde{x}}}{\sqrt{\delta} m} \hspace{0.4cm}, \Phi=\delta \tilde{\Phi} + \frac{\mu}{m}
\end{equation}
The equations of motion become:
\begin{align}
    \label{eqn: dimensionless eqn}
    \tilde{\nabla}^{2} \tilde{\chi} = 2 \left( \tilde{\Phi} \tilde{\chi} - \tilde{\chi^{3}} \right)
    \\
    \tilde{\nabla}^{2} \tilde{\Phi} = \tilde{\chi^{2}}
\end{align}
They satisfy the boundary conditions $\tilde{\chi}'(0)=0, \tilde{\chi}(\tilde{x} \rightarrow \infty)=0, \tilde{\Phi}' (0) = 0$, while the condition $\Phi(\tilde{x} \rightarrow \infty)=0$ implies that $\tilde{\Phi} (\tilde{x} \rightarrow \infty) = -\frac{\mu}{\delta m} $. We see from the last equation that $\frac{\mu}{m} \sim \delta \ll 1$ for the entire range of axion decay constants that we are considering. Therefore, in this non-relativistic limit, we will set $\mu=0$ and $\omega=m$.
\par
Equations \ref{eqn: dimensionless eqn} can be solved with the shooting method: for a given central amplitude of the scalar field $\tilde{\chi_{0}}$, we vary the central amplitude of the potential $\tilde{\Phi_{0}}$ until we find a solution that satisfies the boundary conditions\cite{boson_star_sidm_eby}. Having found the solution, we can also compute the rescaled mass of the axion star $\tilde{M}=\frac{4 \pi G m M}{\sqrt{\delta}}$:
\begin{equation}
    \tilde{M} \approx \int d^{3} \tilde{x} \tilde{\chi}^{2}
\end{equation}
as well as the radius that contains $99 \%$ of its mass. The mass-radius graph generally contains two different branches, the dilute and the transition branch \cite{Wilczek,global_view,boson_star_sidm_eby,Chavanis1}. We are only interested in the dilute branch whose mass-radius graph is depicted in figure \ref{fig:mass_radius} and we confirm that $M \sim \frac{1}{R}$. We also provide the graph of the radius versus the amplitude in figure \ref{fig:radius_ampl} which confirms a well-known behavior of axion stars: as the amplitude increases, the star becomes more dense.

\section{Properties of ISM and radio telescopes}
Before tackling the question of the emitted flux of photons from the axion star, we will consider some properties of the ISM of the Milky Way, such as its large scale magnetic field and two different regions where a resonant conversion may take place, the free electrons and HII regions in nebulae. 

\subsection{Galactic Magnetic field}
\par The magnetic field of the galaxy that we are considering in this work has two components, a large scale one and a small scale one.
\par The large scale component is coherent on length scales of the order of the galaxy and its strength is typically around $1.5 - 2 \mu G$. It reaches $6 \mu G$ in the solar neighborhood and even $10 \mu G$ towards the galactic center. Its structure also seems to follow the spiral arms \cite{B_field_in_milky}. For the purposes of this discussion, the important point is that it is coherent on scales much larger than the size of the axion star and we will consider a value of $1 \mu G$ for its strength. We will say a few things about the small scale component towards the end, but we will ignore it for the remainder of this paper.

\subsection{Free electron density}
\par Regarding the free electron density in the Milky Way disk, that is of the order $0.03 \text{cm}^{-3}$ \cite{radio_astronomy}. Detailed models of the electron density in the Milky Way indicate that the electron density can be as high as $n_{e}=0.2 cm^{-3}$ in the thin disk, while the Local Arm has relatively low density with $n_{e}=0.0057 cm^{-3}$. 
\par The plasma frequency is given by \cite{radio_astronomy}
\begin{equation}
    \label{eqn: plasma_freq}
    \omega_{pl}= 8.97 \text{kHz} \left( \frac{n_{e}}{cm^{-3}} \right)^{1/2} \sim 6 \times 10^{-12} \text{eV}  \left( \frac{n_{e}}{cm^{-3}} \right)^{1/2}
\end{equation}

Given the range of values for the electron density quoted above, the range of plasma frequencies, and therefore axion masses, that we can probe are 0.6 \text{KHz} - 4 \text{kHz} $\rightarrow$ $ 4 \times 10^{-13} \text{eV}$ - $26.3 \times 10^{-13} \text{eV} $. Unfortunately, any electromagnetic waves coming from space with frequency below 30 MHz are blocked by the ionosphere. In addition to that, the solar wind at the Earth's radius can block frequencies that are below 30 $\text{kHz}$. Thus, we conclude that axion stars conversions in environments with free electrons will not produce a detectable flux.

\subsection{HII regions}
We turn to diffuse and planetary nebulae in the interstellar medium with HII regions, that is, ionized hydrogen and electrons. These are formed by stars with temperatures $T \sim 10^{4} K$ that emit UV photons that can ionize the surrounding hydrogen gas, forming the well-known Str\"{o}mgren radius \cite{Stromgren}. The ionization fraction $x=\frac{n_{e}}{n}$, where n is the number density of protons and neutral hydrogen atoms, is equal to unity and the electron densities are generally in the range $100-1000 \text{cm}^{-3}$ \cite{ISM,diffuse_neb}. From equation \ref{eqn: plasma_freq}, these densities correspond to plasma frequencies $90-285 \text{kHz} \Rightarrow 6 \times 10^{-11} - 2 \times 10^{-10} \text{eV}$. There are also ultracompact HII regions that can reach densities $n_{e} \geq 10^{4} \text{cm}^{-3}$ \cite{UCH}, which would correspond to plasma frequencies $\omega_{pl} \geq 897 \text{kHz} = 5.9 \times 10^{-10} \text{eV} $.  We see that the frequency of photons produced in HII regions will be safely above the solar wind's cut-off frequency at the Earth's location. We have ignored here the contribution of protons to the plasma frequency since their mass is much greater than the electron mass.  

\subsection{Radio Telescopes}
One way to observe the low frequencies we are considering here is with lunar or space based telescopes that will not face the issue of the ionosphere. So far, there have been four space or lunar based missions that probed frequencies below 10 MHz \cite{space_radio_Tele}, with the most recent one being the Netherlands Chinese Low Frequency Explorer (NCLE) on board the Chang'e-4 satellite, which has landed on the far side of the Moon and is able to detect frequencies in the range 80 $\text{kHz}$ - 80 $\text{MHz}$ \cite{NCLE}. Hence, it could potentially detect a radio signal from the conversion of an axion star to photons in a HII region. Several other proposals of radio telescopes in space are also described in \cite{space_radio_Tele} that aim to probe the frequency range that is relevant in this paper. 

\section{Spectral flux density}

Having discussed the different environments where the conversion of an axion star to photons may take place, let us now turn to to the calculation of the emitted flux of radio photons that will arrive to Earth when the conversion takes place in a HII region. We will derive an expression for the emitted power from a dilute axion star and demonstate that it can be much larger than the flux from a dense axion star, as well as the flux of the solar system's planets.
\par
To begin, we briefly review the derivation of the emitted flux from an axion star in an external, constant magnetic field. The interaction of the axion with the electromagnetic field is given by the interaction Lagrangian:

\begin{equation}
    \label{eqn: int lagrangian}
    \mathcal{L}_{int}=-\frac{g_{a \gamma}}{4} \phi F_{\mu \nu} \tilde{F}^{\mu \nu}
\end{equation}

We will make a few simplifying assumptions: firstly, we expand the electromagnetic fields and current in the small parameter $g_{a \gamma} \phi_{0}$:

\begin{subequations}
\begin{align}
\bf{E}=\bf{E}^{(0)}+\bf{E}^{(1)}+...,
\qquad
\bf{B}=\bf{B}^{(0)}+\bf{B}^{(1)}+...
\end{align}
\end{subequations}

We also assume zero background electric field and consider the Ansatz of the axion field $\phi(r,t)=\phi_{0} \text{cos}(\omega t) \text{sech}(r/R)$. This choice implies that we will not take back-reaction effects into account, something that we will justify in the next section. The interaction Lagrangian \ref{eqn: int lagrangian} implies the effective current and charge densities $\textbf{J}_{eff}=-g_{a \gamma } \text{sin}(\omega t) \text{sech}(r/R) \textbf{B}_{0}$ and $\rho_{eff}=-g_{a\gamma}\nabla\varphi \cdot \textbf{B}_{0}$. Using the usual definition of electric and magnetic fields in terms of the potentials, we obtain the wave equations: 

\begin{subequations}
\begin{equation}
\label{eqn:wave_equation_potential}
(-\nabla^{2} + \frac{\partial^{2}}{\partial t^{2}} )A^{0} = \rho_{eff} \\
\end{equation}
\begin{equation}
\label{eqn:wave_equation_vector}
(-\nabla^{2} + \frac{\partial^{2}}{ \partial t^{2}} )\textbf{A}  = \textbf{J}_{eff}
\end{equation}
\end{subequations}
where $\textbf{A}$ and $A^{0}$ are first order in $g_{a \gamma} \phi_{0}$. 
\par
In Lorentz gauge, it is enough to solve equation \ref{eqn:wave_equation_vector}, since we can always solve for $A^{0}$ from the equation $\partial_{t}A^{0}+\nabla \cdot \textbf{A}=0$. 
\par
The details of the solution are analyzed in \cite{time_dependent,dipole, Kyriazis} and we will only give the main result here for the emitted power per solid angle: 
\begin{equation}
        \label{eqn: solid angle }
        \frac{dP}{d \Omega}  = \frac{\pi^{4} (g_{a \gamma} \varphi_{0} \omega ^{2} R^{2})^{2}}{32 k \omega} \left( \frac{\tanh (\pi k R /2)}{\cosh (\pi k R /2)} \right)^{2} |\textbf{B}_{0}|^{2}
\end{equation}
where k is the wave-vector of the emitted photons and it is equal to $k=\omega \sqrt{1-\frac{\omega_{p}^{2}}{\omega^{2}}}$, if we also include cold plasma. We have implicitly assumed that the gyrofrequency $\omega_{B}=\frac{\sqrt{4 \pi \alpha} B_{0}}{m_{e}}$ is much smaller than the frequency of radiation $\omega$, which is true for the values we are considering here: $\omega_{B}=10^{-15} \text{eV} \ll \omega=10^{-10} \text{eV} $. One final assumption is that the propagation of the photons is perpendicular to the galactic magnetic field, because we are mainly interested in an order of magnitude estimate of the effect. 

\begin{figure}[t!]
    \centering
        \includegraphics[width=1\textwidth]{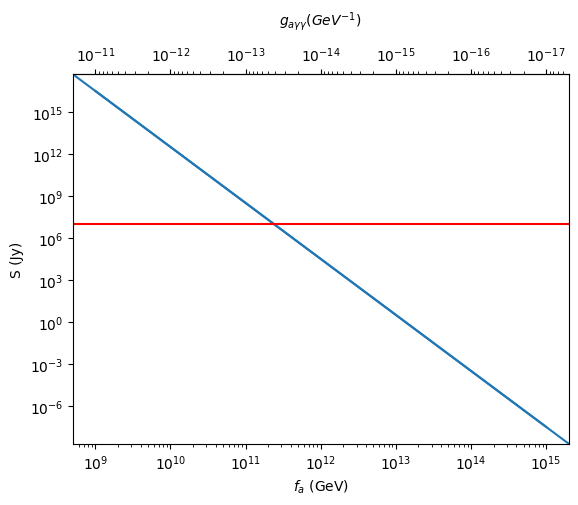}
        \caption{Spectral flux density in Jy versus the axion decay constant for electron density 400 $\text{cm}^{-3}$. The red horizontal line is at $10^{7}$ Jy, approximately the radio emission from Saturn in this frequency range}
        \label{fig:flux_f}
\end{figure}

\par
We see that when the axion star is far from resonance, the power peaks for sizes $ \omega R \sim 1$ while it is exponentially suppressed if $\omega R \gg 1$. We are interested in studying the power emitted when the star is in resonance, so we take $ \omega \rightarrow \omega_{p} \sim 10^{-10} \text{eV} $. The power becomes:
\begin{equation}
    \frac{dP}{d\Omega}({\omega \rightarrow \omega_{p}}) \approx  \frac{(g_{a \gamma} \phi_{0})^{2}}{128} \frac{(\pi \omega R)^{6}}{\omega^{2}} \left(1-\frac{\omega_{p}^{2}}{\omega^{2}}\right)^{1/2} |\textbf{B}_{0}|^{2}
\end{equation}

In the spirit of the order of magnitude estimate that we are attempting, let us assume that this conversion takes place 1 kpc away from Earth, which is a typical galactic distance. We define the spectral flux density of the incoming radiation as $S=\frac{1}{r^{2} \mathcal{B}} \frac{dP}{d \Omega}$ where $\mathcal{B}$ is the Doppler shift of the central frequency and we estimate it as $\mathcal{B} \sim \frac{0.1\omega}{2 \pi}$. For a dense axion star, the value of the spectral flux density is of the order: $S \sim 10^{-21} Jy$, which is indeed negligible. 

\par However, the $(\omega R)^{6}$ term is promising: dilute axion stars with $\omega R \gg 1$ can significantly enhance the flux that arrives at Earth. To find the power emitted from a dilute axion star in resonance, we make the substitution $\phi_{0} \rightarrow \sqrt{\frac{1}{2 \pi G}}  \delta$, which comes from combining equations \ref{eqn:rescale} and \ref{eqn: Ansatz}. Assuming that $g_{a \gamma} \sim \frac{\alpha}{f_{a}}$ and with the approximation $\omega \approx m$, our estimate is:

\begin{equation}
    S \sim 10^{-21} \text{Jy} \frac{1}{\delta^{2}}  \left( \frac{1 kpc}{r} \right)^{2} \left( \frac{10^{-10} \text{eV}}{m} \right)^{3} \left( \frac{B_{0}}{1 \mu G} \right)^{2} \
\end{equation}

where we have also approximated $mR \sim \frac{1}{\sqrt{\delta}}$. We see that since $\delta \ll 1$ for a wide range of decay constants, this flux can be quite huge.
\par
Our plan is to compare this flux density with the fluxes that come from the planets of our solar system. In the frequency range 100-300 kHz, the largest flux comes from Saturn which reaches approximately $10^{-19} \frac{\text{W}}{\text{Hz} \hspace{0.06cm} \text{m}^{2}}=10^{7} \text{Jy}$ \cite{NOIRE}. 
\par
The spectral flux density for different axion decay constants and electron density set to $400 \text{cm}^{-3}$ is depicted in figure \ref{fig:flux_f}. The red horizontal line corresponds to the flux coming from Saturn. We see that for axion decay constants approximately smaller than $5 \times 10^{11} \text{GeV}$  the signal is above Saturn's threshold and therefore it could be detected by a space or lunar based radio telescope that is sensitive to the corresponding frequency range.  

\section{Decay time scales} \label{sec:decay}
We estimate in this section the time that it will take for the axion star to convert all its mass to photons. The mass of a dilute axion star is of the order of $M \sim 10 \frac{f_{a} M_{pl}}{m}$ \cite{Wilczek}. We assume the star is in resonance with the surrounding plasma. The timescale over which the star will lose the entirety of this mass is roughly given by:
\begin{equation}
    T=\frac{M}{P} \sim 10^{4} \frac{M^{2}_{pl} m \delta^{5/2}}{B_{0}^{2}}
\end{equation}
where we have estimated the emitted power to be $P \sim 10^{-3} \frac{1}{\delta^{2}} \frac{B^{2}_{0}}{m^{2}}$. We are ignoring back-reaction effects here, so in order for this estimation to make sense, we need $T> \frac{2 \pi}{m}$, the decay time scale needs to be longer than the period of the radiation \cite{time_dependent}.Solving for the small parameter $\delta$, we find 
\begin{equation}
    \delta> \left( \frac{10^{-2} B_{0}}{m M_{pl}} \right)^{4/5} \Rightarrow f_{a} > M_{pl} \left( \frac{10^{-2} B_{0}}{m M_{pl}} \right)^{2/5}
\end{equation}
For $B_{0}= 1 \mu \text{G}$ and $m=10^{-10} \text{eV}$, this gives the lower bound  $f_{a} \gtrapprox 10^{7} \text{GeV}$. This tells us that for most of the parameter space, back-reaction can indeed be ignored. For $f_{a}=10^{9} \text{GeV}$, the lower bound on the decay timescale is $T \sim 3 \text{hr}$.  

\section{Conclusions and Outlook}
\par
We have considered the possibility of axion stars converting to photons in the magnetic field of the Milky Way. The high number of axion stars in the galactic disk that we estimated in the Introduction make this a possibility worth considering. We showed that the weak magnetic field of the galaxy is not enough to efficiently convert a dense axion star to photons in vacuum. However, if an axion star is in a plasma and its frequency is close to the plasma frequency, the dependence of the emitted flux on $(mR)^{6}$ implies that a dilute axion star will produce a sizable flux. 

\par Beginning from this observation, we considered two different possibilities of a plasma in the Milky Way, the free electrons and the diffuse nebulae with HII regions. In the former case, we argued that the photons produced will have frequencies far below the solar wind cut-off and we will never be able to observe them with lunar based radio-telescopes. In the latter case, however, the electron density is $10^{4}$ time larger, leading to photons with frequencies $\nu \geq 100 \text{kHz}$ safely above the solar wind threshold and within the target range of current lunar based telescopes, such as NCLE.

\par Our main calculation involved the estimation of the spectral flux density that will arrive at Earth if a dilute axion star resonated with its surrounding plasma and converted its mass to photons. We showed that for axion decay constants $f_{a}< 5 \times 10^{11} \text{GeV}$, the flux is larger than the radio flux emitted from Saturn, which is the dominant one from the solar system's planets in this frequency range.

\par Finally, we estimated the time scale over which the star will radiate. We demonstrated that back-reaction effects can be ignored for the entirety of the parameter space that we consider in this work and found that an axion star will need at least a few hours to lose all its mass. However, it is an open question whether the star will transition to different configurations as it decays, so that it eventually moves out of resonance. Ref. \cite{axion_backreaction} did this analysis for dense axion stars supported by their self-interactions and found that the axion stars grow in size, their frequency increases and they go out of resonance after a certain time-scale. It is not clear whether the same thing can happen with dilute stars because, to a very good approximation, $\omega \sim m$. An analysis in the vein of \cite{Levkov} could shed some light on this question. 

\par A coulpe of more comments are in order regarding this proposal. Firstly, we have ignored the small scale component of the galactic field that is related to the turbulent Interstellar Medium . This component has a shorter correlation length than the large-scale component we used in this study and its strength is $5.5 \mu G$ \cite{B_field_in_milky}. It has been shown that it can enhance the conversion of diffuse axions to photons by many orders of magnitude and it should be taken into account in future work\cite{relat_axions_in_sky,turbulent}. 

\par Also, we have not considered the distribution of axion stars in the galactic plane, which, to our knowledge, is not known. This makes it difficult to estimate the number of conversions that we could potentially observe. A more detailed study should take the axion star distribution into account, combined with the disribution of HII in the Milky Way, as shown in \cite{Paladini}. This should provide us with an accurate estimation of the frequency of these events.

\acknowledgments
I would like to thank Yuxin Zhao, Pierre Sikivie and Joshua Eby for useful comments and discussions. This scientific paper was supported by the Onassis Foundation - Scholarship ID: F ZS 031-1/2022-2023.

\bibliography{cit}

\begin{thebibliography}{55}
\providecommand{\natexlab}[1]{#1}
\providecommand{\url}[1]{\texttt{#1}}
\expandafter\ifx\csname urlstyle\endcsname\relax
  \providecommand{\doi}[1]{doi: #1}\else
  \providecommand{\doi}{doi: \begingroup \urlstyle{rm}\Url}\fi

\bibitem[Peccei and Quinn(1977)]{Peccei}
R.~D. Peccei and Helen~R. Quinn.
\newblock $\mathrm{CP}$ conservation in the presence of pseudoparticles.
\newblock \emph{Phys. Rev. Lett.}, 38:\penalty0 1440--1443, Jun 1977.
\newblock \doi{10.1103/PhysRevLett.38.1440}.
\newblock URL \url{https://link.aps.org/doi/10.1103/PhysRevLett.38.1440}.

\bibitem[Weinberg(1978)]{Weinberg}
Steven Weinberg.
\newblock A new light boson?
\newblock \emph{Phys. Rev. Lett.}, 40:\penalty0 223--226, Jan 1978.
\newblock \doi{10.1103/PhysRevLett.40.223}.
\newblock URL \url{https://link.aps.org/doi/10.1103/PhysRevLett.40.223}.

\bibitem[Preskill et~al.(1983)Preskill, Wise, and Wilczek]{Preskill}
John Preskill, Mark~B. Wise, and Frank Wilczek.
\newblock Cosmology of the invisible axion.
\newblock \emph{Physics Letters B}, 120\penalty0 (1):\penalty0 127--132, 1983.
\newblock ISSN 0370-2693.
\newblock \doi{https://doi.org/10.1016/0370-2693(83)90637-8}.
\newblock URL
  \url{https://www.sciencedirect.com/science/article/pii/0370269383906378}.

\bibitem[Abbott and Sikivie(1983)]{Sikivie}
L.F. Abbott and P.~Sikivie.
\newblock A cosmological bound on the invisible axion.
\newblock \emph{Physics Letters B}, 120\penalty0 (1):\penalty0 133--136, 1983.
\newblock ISSN 0370-2693.
\newblock \doi{https://doi.org/10.1016/0370-2693(83)90638-X}.
\newblock URL
  \url{https://www.sciencedirect.com/science/article/pii/037026938390638X}.

\bibitem[Dine and Fischler(1983)]{Willy}
Michael Dine and Willy Fischler.
\newblock The not-so-harmless axion.
\newblock \emph{Physics Letters B}, 120\penalty0 (1):\penalty0 137--141, 1983.
\newblock ISSN 0370-2693.
\newblock \doi{https://doi.org/10.1016/0370-2693(83)90639-1}.
\newblock URL
  \url{https://www.sciencedirect.com/science/article/pii/0370269383906391}.

\bibitem[Sikivie and Yang(2009)]{BEC1}
P.~Sikivie and Q.~Yang.
\newblock Bose-einstein condensation of dark matter axions.
\newblock \emph{Physical Review Letters}, 103\penalty0 (11), sep 2009.
\newblock \doi{10.1103/physrevlett.103.111301}.
\newblock URL \url{https://doi.org/10.1103%2Fphysrevlett.103.111301}.

\bibitem[Erken et~al.(2012)Erken, Sikivie, Tam, and Yang]{BEC2}
O.~Erken, P.~Sikivie, H.~Tam, and Q.~Yang.
\newblock Axion dark matter and cosmological parameters.
\newblock \emph{Physical Review Letters}, 108\penalty0 (6), feb 2012.
\newblock \doi{10.1103/physrevlett.108.061304}.
\newblock URL \url{https://doi.org/10.1103%2Fphysrevlett.108.061304}.

\bibitem[Braaten et~al.(2016)Braaten, Mohapatra, and Zhang]{axion_star1}
Eric Braaten, Abhishek Mohapatra, and Hong Zhang.
\newblock Dense axion stars.
\newblock \emph{Phys. Rev. Lett.}, 117:\penalty0 121801, Sep 2016.
\newblock \doi{10.1103/PhysRevLett.117.121801}.
\newblock URL \url{https://link.aps.org/doi/10.1103/PhysRevLett.117.121801}.

\bibitem[Zhang et~al.(2020)Zhang, Amin, Copeland, Saffin, and Lozanov]{decay}
Hong-Yi Zhang, Mustafa~A. Amin, Edmund~J. Copeland, Paul~M. Saffin, and
  Kaloian~D. Lozanov.
\newblock Classical decay rates of oscillons.
\newblock \emph{Journal of Cosmology and Astroparticle Physics}, 2020\penalty0
  (07):\penalty0 055--055, jul 2020.
\newblock \doi{10.1088/1475-7516/2020/07/055}.
\newblock URL \url{https://doi.org/10.1088/1475-7516/2020/07/055}.

\bibitem[Visinelli et~al.(2018)Visinelli, Baum, Redondo, Freese, and
  Wilczek]{Wilczek}
Luca Visinelli, Sebastian Baum, Javier Redondo, Katherine Freese, and Frank
  Wilczek.
\newblock Dilute and dense axion stars.
\newblock \emph{Physics Letters B}, 777:\penalty0 64--72, 2018.
\newblock ISSN 0370-2693.
\newblock \doi{https://doi.org/10.1016/j.physletb.2017.12.010}.
\newblock URL
  \url{https://www.sciencedirect.com/science/article/pii/S0370269317309875}.

\bibitem[Chavanis(2018)]{Chavanis}
Pierre-Henri Chavanis.
\newblock Phase transitions between dilute and dense axion stars.
\newblock \emph{Physical Review D}, 98\penalty0 (2), jul 2018.
\newblock \doi{10.1103/physrevd.98.023009}.
\newblock URL \url{https://doi.org/10.1103%2Fphysrevd.98.023009}.

\bibitem[Marsh(2016)]{axion_cosmology}
David~J.E. Marsh.
\newblock Axion cosmology.
\newblock \emph{Physics Reports}, 643:\penalty0 1--79, jul 2016.
\newblock \doi{10.1016/j.physrep.2016.06.005}.
\newblock URL \url{https://doi.org/10.1016%2Fj.physrep.2016.06.005}.

\bibitem[Eby et~al.(2016{\natexlab{a}})Eby, Suranyi, and
  Wijewardhana]{axion_star2}
Joshua Eby, Peter Suranyi, and L.~C.~R. Wijewardhana.
\newblock The lifetime of axion stars.
\newblock \emph{Modern Physics Letters A}, 31\penalty0 (15):\penalty0 1650090,
  may 2016{\natexlab{a}}.
\newblock \doi{10.1142/s0217732316500905}.
\newblock URL \url{https://doi.org/10.1142%2Fs0217732316500905}.

\bibitem[Eby et~al.(2021)Eby, Street, Suranyi, and Wijewardhana]{planck_decay}
Joshua Eby, Lauren Street, Peter Suranyi, and L.{\hspace{0.167em}
  }C.{\hspace{0.167em}}R. Wijewardhana.
\newblock Global view of axion stars with nearly planck-scale decay constants.
\newblock \emph{Physical Review D}, 103\penalty0 (6), mar 2021.
\newblock \doi{10.1103/physrevd.103.063043}.
\newblock URL \url{https://doi.org/10.1103%2Fphysrevd.103.063043}.

\bibitem[Sikivie(1983)]{axion_detection}
P.~Sikivie.
\newblock Experimental tests of the "invisible" axion.
\newblock \emph{Phys. Rev. Lett.}, 51:\penalty0 1415--1417, Oct 1983.
\newblock \doi{10.1103/PhysRevLett.51.1415}.
\newblock URL \url{https://link.aps.org/doi/10.1103/PhysRevLett.51.1415}.

\bibitem[Primakoff(1951)]{Primakoff}
H.~Primakoff.
\newblock Photo-production of neutral mesons in nuclear electric fields and the
  mean life of the neutral meson.
\newblock \emph{Phys. Rev.}, 81:\penalty0 899--899, Mar 1951.
\newblock \doi{10.1103/PhysRev.81.899}.
\newblock URL \url{https://link.aps.org/doi/10.1103/PhysRev.81.899}.

\bibitem[Pshirkov and Popov(2009)]{axion_neutron}
M.~S. Pshirkov and S.~B. Popov.
\newblock Conversion of dark matter axions to photons in magnetospheres of
  neutron stars.
\newblock \emph{Journal of Experimental and Theoretical Physics}, 108\penalty0
  (3):\penalty0 384--388, mar 2009.
\newblock \doi{10.1134/s1063776109030030}.
\newblock URL \url{https://doi.org/10.1134%2Fs1063776109030030}.

\bibitem[Bai and Hamada(2018)]{Bai}
Yang Bai and Yuta Hamada.
\newblock Detecting axion stars with radio telescopes.
\newblock \emph{Physics Letters B}, 781:\penalty0 187--194, jun 2018.
\newblock \doi{10.1016/j.physletb.2018.03.070}.
\newblock URL \url{https://doi.org/10.1016%2Fj.physletb.2018.03.070}.

\bibitem[Witte et~al.(2023)Witte, Baum, Lawson, Marsh, Millar, and
  Salinas]{transient_radio}
Samuel~J. Witte, Sebastian Baum, Matthew Lawson, M.~C.~David Marsh,
  Alexander~J. Millar, and Gustavo Salinas.
\newblock Transient radio lines from axion miniclusters and axion stars.
\newblock \emph{Phys. Rev. D}, 107:\penalty0 063013, Mar 2023.
\newblock \doi{10.1103/PhysRevD.107.063013}.
\newblock URL \url{https://link.aps.org/doi/10.1103/PhysRevD.107.063013}.

\bibitem[Hook et~al.(2018)Hook, Kahn, Safdi, and Sun]{Radio_signals}
Anson Hook, Yonatan Kahn, Benjamin~R. Safdi, and Zhiquan Sun.
\newblock Radio signals from axion dark matter conversion in neutron star
  magnetospheres.
\newblock \emph{Phys. Rev. Lett.}, 121:\penalty0 241102, Dec 2018.
\newblock \doi{10.1103/PhysRevLett.121.241102}.
\newblock URL \url{https://link.aps.org/doi/10.1103/PhysRevLett.121.241102}.

\bibitem[Battye et~al.(2021)Battye, Garbrecht, McDonald, and
  Srinivasan]{Radio_line}
R.~A. Battye, B.~Garbrecht, J.~McDonald, and S.~Srinivasan.
\newblock Radio line properties of axion dark matter conversion in neutron
  stars.
\newblock \emph{Journal of High Energy Physics}, 2021\penalty0 (9), sep 2021.
\newblock \doi{10.1007/jhep09(2021)105}.
\newblock URL \url{https://doi.org/10.1007%2Fjhep09%282021%29105}.

\bibitem[Buckley et~al.(2021)Buckley, Dev, Ferrer, and Huang]{FRB}
James~H. Buckley, P.{\hspace{0.167em} }S.~Bhupal Dev, Francesc Ferrer, and
  Fa~Peng Huang.
\newblock Fast radio bursts from axion stars moving through pulsar
  magnetospheres.
\newblock \emph{Physical Review D}, 103\penalty0 (4), feb 2021.
\newblock \doi{10.1103/physrevd.103.043015}.
\newblock URL \url{https://doi.org/10.1103%2Fphysrevd.103.043015}.

\bibitem[Fortin and Sinha(2019)]{magnetar1}
Jean-Fran{\c{c}}ois Fortin and Kuver Sinha.
\newblock X-ray polarization signals from magnetars with axion-like-particles.
\newblock \emph{Journal of High Energy Physics}, 2019\penalty0 (1), jan 2019.
\newblock \doi{10.1007/jhep01(2019)163}.
\newblock URL \url{https://doi.org/10.1007%2Fjhep01%282019%29163}.

\bibitem[Fortin and Sinha(2018)]{magnetar2}
Jean-Fran{\c{c}}ois Fortin and Kuver Sinha.
\newblock Constraining axion-like-particles with hard x-ray emission from
  magnetars.
\newblock \emph{Journal of High Energy Physics}, 2018\penalty0 (6), jun 2018.
\newblock \doi{10.1007/jhep06(2018)048}.
\newblock URL \url{https://doi.org/10.1007%2Fjhep06%282018%29048}.

\bibitem[Fortin et~al.(2021)Fortin, Guo, Harris, Sheridan, and
  Sinha]{magnetar3}
Jean-Fran{\c{c} }ois Fortin, Huai-Ke Guo, Steven~P. Harris, Elijah Sheridan,
  and Kuver Sinha.
\newblock Magnetars and axion-like particles: probes with the hard x-ray
  spectrum.
\newblock \emph{Journal of Cosmology and Astroparticle Physics}, 2021\penalty0
  (06):\penalty0 036, jun 2021.
\newblock \doi{10.1088/1475-7516/2021/06/036}.
\newblock URL \url{https://doi.org/10.1088%2F1475-7516%2F2021%2F06%2F036}.

\bibitem[Dessert et~al.(2022)Dessert, Dunsky, and
  Safdi]{upper_limit_from_axion_photon}
Christopher Dessert, David Dunsky, and Benjamin~R. Safdi.
\newblock Upper limit on the axion-photon coupling from magnetic white dwarf
  polarization.
\newblock \emph{Physical Review D}, 105\penalty0 (10), may 2022.
\newblock \doi{10.1103/physrevd.105.103034}.
\newblock URL \url{https://doi.org/10.1103%2Fphysrevd.105.103034}.

\bibitem[Balkin et~al.(2022)Balkin, Serra, Springmann, Stelzl, and
  Weiler]{white_dwarfs}
Reuven Balkin, Javi Serra, Konstantin Springmann, Stefan Stelzl, and Andreas
  Weiler.
\newblock White dwarfs as a probe of light qcd axions, 2022.

\bibitem[Chang et~al.(2023)Chang, Ebadi, Luo, and Tanin]{spectral_distortions}
Jae~Hyeok Chang, Reza Ebadi, Xuheng Luo, and Erwin~H. Tanin.
\newblock Spectral distortions of astrophysical blackbodies as axion probes,
  2023.

\bibitem[Ayad and Beck(2020)]{AGN1}
Ahmed Ayad and Geoff Beck.
\newblock Probing a cosmic axion-like particle background within the jets of
  active galactic nuclei.
\newblock \emph{Journal of Cosmology and Astroparticle Physics}, 2020\penalty0
  (04):\penalty0 055--055, apr 2020.
\newblock \doi{10.1088/1475-7516/2020/04/055}.
\newblock URL \url{https://doi.org/10.1088%2F1475-7516%2F2020%2F04%2F055}.

\bibitem[Harris and Chadwick(2014)]{AGN2}
J.~Harris and P.M. Chadwick.
\newblock Photon-axion mixing within the jets of active galactic nuclei and
  prospects for detection.
\newblock \emph{Journal of Cosmology and Astroparticle Physics}, 2014\penalty0
  (10):\penalty0 018--018, oct 2014.
\newblock \doi{10.1088/1475-7516/2014/10/018}.
\newblock URL \url{https://doi.org/10.1088%2F1475-7516%2F2014%2F10%2F018}.

\bibitem[Carenza et~al.(2021)Carenza, Evoli, Giannotti, Mirizzi, and
  Montanino]{turbulent}
Pierluca Carenza, Carmelo Evoli, Maurizio Giannotti, Alessandro Mirizzi, and
  Daniele Montanino.
\newblock Turbulent axion-photon conversions in the milky~way.
\newblock \emph{Physical Review D}, 104\penalty0 (2), jul 2021.
\newblock \doi{10.1103/physrevd.104.023003}.
\newblock URL \url{https://doi.org/10.1103%2Fphysrevd.104.023003}.

\bibitem[Kelley and Quinn(2017)]{axion_dm_radio_tele}
Katharine Kelley and P.~J. Quinn.
\newblock Searching for axion dark matter using radio telescopes, 2017.

\bibitem[{Kar} et~al.(2022){Kar}, {Kumar}, {Roy}, and
  {Zupan}]{relat_axions_in_sky}
Arpan {Kar}, Tanmoy {Kumar}, Sourov {Roy}, and Jure {Zupan}.
\newblock {Searching for relativistic axions in the sky}.
\newblock \emph{arXiv e-prints}, December 2022.
\newblock \doi{10.48550/arXiv.2212.04647}.

\bibitem[Fairbairn(2014)]{axionic_radiation}
Malcolm Fairbairn.
\newblock Axionic dark radiation and the milky~way's magnetic field.
\newblock \emph{Physical Review D}, 89\penalty0 (6), mar 2014.
\newblock \doi{10.1103/physrevd.89.064020}.
\newblock URL \url{https://doi.org/10.1103%2Fphysrevd.89.064020}.

\bibitem[Haverkorn(2014)]{B_field_in_milky}
Marijke Haverkorn.
\newblock Magnetic fields in the milky way.
\newblock In \emph{Astrophysics and Space Science Library}, pages 483--506.
  Springer Berlin Heidelberg, oct 2014.
\newblock \doi{10.1007/978-3-662-44625-6_17}.
\newblock URL \url{https://doi.org/10.1007%2F978-3-662-44625-6_17}.

\bibitem[Amin et~al.(2021)Amin, Long, Mou, and Saffin]{dipole}
Mustafa~A. Amin, Andrew~J. Long, Zong-Gang Mou, and Paul~M. Saffin.
\newblock Dipole radiation and beyond from axion stars in electromagnetic
  fields.
\newblock \emph{Journal of High Energy Physics}, 2021\penalty0 (6):\penalty0
  182, 2021.
\newblock ISSN 1029-8479.
\newblock \doi{10.1007/JHEP06(2021)182}.
\newblock URL \url{https://doi.org/10.1007/JHEP06(2021)182}.

\bibitem[Sen and Sivertsen(2021)]{time_dependent}
Srimoyee Sen and Lars Sivertsen.
\newblock Electromagnetic radiation from axion condensates in a time dependent
  magnetic field, 2021.
\newblock URL \url{https://arxiv.org/abs/2111.08728}.

\bibitem[{Condon} and {Ransom}(2016)]{radio_astronomy}
James~J. {Condon} and Scott~M. {Ransom}.
\newblock \emph{{Essential Radio Astronomy}}.
\newblock 2016.

\bibitem[{Kwok}(2007)]{ISM}
Sun {Kwok}.
\newblock \emph{{Physics and Chemistry of the Interstellar Medium}}.
\newblock 2007.

\bibitem[Prinsloo et~al.(2018)Prinsloo, Ruiter, Arts, Marel, Boonstra,
  Kruithof, Wise, Falcke, Klein-Wolt, Rothkaehl, Cecconi, Dekkali, and
  Ping]{NCLE}
D.S. Prinsloo, M.~Ruiter, M.J. Arts, J.~V.~D. Marel, A.J. Boonstra,
  G.~Kruithof, M.~Wise, H.~Falcke, M.~Klein-Wolt, H.~Rothkaehl, B.~Cecconi,
  M.~Dekkali, and J.~Ping.
\newblock Emi modelling of an 80 khz to 80 mhz wideband antenna and low-noise
  amplifier for radio astronomy in space.
\newblock In \emph{12th European Conference on Antennas and Propagation (EuCAP
  2018)}, pages 1--4, 2018.
\newblock \doi{10.1049/cp.2018.0820}.

\bibitem[Arvanitaki et~al.(2010)Arvanitaki, Dimopoulos, Dubovsky, Kaloper, and
  March-Russell]{ALP}
Asimina Arvanitaki, Savas Dimopoulos, Sergei Dubovsky, Nemanja Kaloper, and
  John March-Russell.
\newblock String axiverse.
\newblock \emph{Physical Review D}, 81\penalty0 (12), jun 2010.
\newblock \doi{10.1103/physrevd.81.123530}.
\newblock URL \url{https://doi.org/10.1103%2Fphysrevd.81.123530}.

\bibitem[Kouvaris et~al.(2022)Kouvaris, Liu, and Lyu]{ANS_binaries}
Chris Kouvaris, Tao Liu, and Kun-Feng Lyu.
\newblock Radio signals from axion star-neutron star binaries, 2022.
\newblock URL \url{https://arxiv.org/abs/2202.11096}.

\bibitem[Banik et~al.(2015)Banik, Christopherson, Sikivie, and
  Todarello]{linear_newt}
Nilanjan Banik, Adam~J. Christopherson, Pierre Sikivie, and Elisa~Maria
  Todarello.
\newblock Linear newtonian perturbation theory from the schr\"odinger-poisson
  equations, 2015.

\bibitem[Eby et~al.(2019)Eby, Leembruggen, Street, Suranyi, and
  Wijewardhana]{global_view}
Joshua Eby, Madelyn Leembruggen, Lauren Street, Peter Suranyi, and
  L.{\hspace{0.167em} }C.{\hspace{0.167em}}R. Wijewardhana.
\newblock Global view of {QCD} axion stars.
\newblock \emph{Physical Review D}, 100\penalty0 (6), sep 2019.
\newblock \doi{10.1103/physrevd.100.063002}.
\newblock URL \url{https://doi.org/10.1103%2Fphysrevd.100.063002}.

\bibitem[Eby et~al.(2016{\natexlab{b}})Eby, Kouvaris, Nielsen, and
  Wijewardhana]{boson_star_sidm_eby}
Joshua Eby, Chris Kouvaris, Niklas~Gr{\o}nlund Nielsen, and L.~C.~R.
  Wijewardhana.
\newblock Boson stars from self-interacting dark matter.
\newblock \emph{Journal of High Energy Physics}, 2016\penalty0 (2), feb
  2016{\natexlab{b}}.
\newblock \doi{10.1007/jhep02(2016)028}.
\newblock URL \url{https://doi.org/10.1007%2Fjhep02%282016%29028}.

\bibitem[Chavanis(2011)]{Chavanis1}
Pierre-Henri Chavanis.
\newblock Mass-radius relation of newtonian self-gravitating bose-einstein
  condensates with short-range interactions. i. analytical results.
\newblock \emph{Physical Review D}, 84\penalty0 (4), aug 2011.
\newblock \doi{10.1103/physrevd.84.043531}.
\newblock URL \url{https://doi.org/10.1103%2Fphysrevd.84.043531}.

\bibitem[{Str{\"o}mgren}(1939)]{Stromgren}
Bengt {Str{\"o}mgren}.
\newblock {The Physical State of Interstellar Hydrogen.}
\newblock \emph{\apj}, 89:\penalty0 526, May 1939.
\newblock \doi{10.1086/144074}.

\bibitem[{Osterbrock}(1967)]{diffuse_neb}
Donald~E. {Osterbrock}.
\newblock {Diffuse Nebulae}.
\newblock \emph{Publications of the Astronomical Society of the Pacific},
  79\penalty0 (471):\penalty0 523, December 1967.
\newblock \doi{10.1086/128550}.

\bibitem[{Hoare} et~al.(2007){Hoare}, {Kurtz}, {Lizano}, {Keto}, and
  {Hofner}]{UCH}
M.~G. {Hoare}, S.~E. {Kurtz}, S.~{Lizano}, E.~{Keto}, and P.~{Hofner}.
\newblock {Ultracompact Hii Regions and the Early Lives of Massive Stars}.
\newblock In Bo~{Reipurth}, David {Jewitt}, and Klaus {Keil}, editors,
  \emph{Protostars and Planets V}, page 181, January 2007.
\newblock \doi{10.48550/arXiv.astro-ph/0603560}.

\bibitem[Bentum et~al.(2020)Bentum, Verma, Rajan, Boonstra, Verhoeven, Gill,
  van~der Veen, Falcke, Wolt, Monna, Engelen, Rotteveel, and
  Gurvits]{space_radio_Tele}
M.J. Bentum, M.K. Verma, R.T. Rajan, A.J. Boonstra, C.J.M. Verhoeven, E.K.A.
  Gill, A.J. van~der Veen, H.~Falcke, M.~Klein Wolt, B.~Monna, S.~Engelen,
  J.~Rotteveel, and L.I. Gurvits.
\newblock A roadmap towards a space-based radio telescope for ultra-low
  frequency radio astronomy.
\newblock \emph{Advances in Space Research}, 65\penalty0 (2):\penalty0
  856--867, jan 2020.
\newblock \doi{10.1016/j.asr.2019.09.007}.
\newblock URL \url{https://doi.org/10.1016%2Fj.asr.2019.09.007}.

\bibitem[Kyriazis(2022)]{Kyriazis}
A.~Kyriazis.
\newblock Radiation from axion star-neutron star binaries with a tilted
  rotation axis in the presence of plasma.
\newblock \emph{Journal of High Energy Physics}, 2022\penalty0 (11), nov 2022.
\newblock \doi{10.1007/jhep11(2022)014}.
\newblock URL \url{https://doi.org/10.1007%2Fjhep11%282022%29014}.

\bibitem[Cecconi et~al.(2018)Cecconi, Dekkali, Briand, Segret, Girard, Laurens,
  Lamy, Valat, Delpech, Bruno, Gelard, Bucher, Nenon, Griesmeier, Boonstra, and
  Bentum]{NOIRE}
Baptiste Cecconi, Moustapha Dekkali, Carine Briand, Boris Segret, Julien~N.
  Girard, Andre Laurens, Alain Lamy, David Valat, Michel Delpech, Mickael
  Bruno, Patrick Gelard, Martin Bucher, Quentin Nenon, Jean-Mathias Griesmeier,
  Albert-Jan Boonstra, and Mark Bentum.
\newblock {NOIRE} study report: Towards a low frequency radio interferometer in
  space.
\newblock In \emph{2018 {IEEE} Aerospace Conference}. {IEEE}, mar 2018.
\newblock \doi{10.1109/aero.2018.8396742}.
\newblock URL \url{https://doi.org/10.1109%2Faero.2018.8396742}.

\bibitem[Sen and Sivertsen(2023)]{axion_backreaction}
Srimoyee Sen and Lars Sivertsen.
\newblock Energy conservation and axion back-reaction in a magnetic field.
\newblock \emph{Journal of High Energy Physics}, 2023\penalty0 (3), mar 2023.
\newblock \doi{10.1007/jhep03(2023)097}.
\newblock URL \url{https://doi.org/10.1007%2Fjhep03%282023%29097}.

\bibitem[Levkov et~al.(2020)Levkov, Panin, and Tkachev]{Levkov}
D.{\hspace{0.167em} }G. Levkov, A.{\hspace{0.167em}}G. Panin, and
  I.{\hspace{0.167em}}I. Tkachev.
\newblock Radio-emission of axion stars.
\newblock \emph{Physical Review D}, 102\penalty0 (2), jul 2020.
\newblock \doi{10.1103/physrevd.102.023501}.
\newblock URL \url{https://doi.org/10.1103%2Fphysrevd.102.023501}.

\bibitem[Paladini et~al.(2004)Paladini, Davies, and DeZotti]{Paladini}
R.~Paladini, R.~D. Davies, and G.~DeZotti.
\newblock Spatial distribution of galactic h{\hspace{1em} }ii regions.
\newblock \emph{Monthly Notices of the Royal Astronomical Society},
  347\penalty0 (1):\penalty0 237--245, jan 2004.
\newblock \doi{10.1111/j.1365-2966.2004.07210.x}.
\newblock URL \url{https://doi.org/10.1111%2Fj.1365-2966.2004.07210.x}.

\end{thebibliography}

\end{document}